\documentclass[twocolumn,english,secnumarabic,eqsecnum]{revtex4}
\usepackage[T1]{fontenc}
\usepackage[latin9]{inputenc}
\usepackage{float}
\usepackage{amsmath}
\usepackage{amssymb}
\usepackage{graphicx}
\usepackage{esint}

\makeatletter
\@ifundefined{textcolor}{}
{%
 \definecolor{BLACK}{gray}{0}
 \definecolor{WHITE}{gray}{1}
 \definecolor{RED}{rgb}{1,0,0}
 \definecolor{GREEN}{rgb}{0,1,0}
 \definecolor{BLUE}{rgb}{0,0,1}
 \definecolor{CYAN}{cmyk}{1,0,0,0}
 \definecolor{MAGENTA}{cmyk}{0,1,0,0}
 \definecolor{YELLOW}{cmyk}{0,0,1,0}
}

\usepackage{babel}

\usepackage{babel}

\usepackage{babel}

\makeatother

\usepackage{babel}
\begin{document}

\title{Two interacting fermions in a 1D harmonic trap: matching the Bethe
ansatz and variational approaches}

\author{D. Rubeni$^{1}$, A. Foerster$^{1}$ and I. Roditi$^{2}$}

\address{$^{1}$Instituto de F\'{i}sica - UFRGS, Porto Alegre, RS - Brazil\\
 $^{2}$Centro Brasileiro de Pesquisas F\'{i}sicas - CBPF, Rio de
Janeiro, RJ - Brazil}
\begin{abstract}
In this work, combining the Bethe ansatz approach with the variational
principle, we calculate the ground state energy of the relative motion
of a system of two fermions with spin up and down interacting via
a delta-function potential in a 1D harmonic trap. Our results show
good agreement with the analytical solution of the problem, and provide
a starting point for the investigation of more complex few-body systems
where no exact theoretical solution is available. 
\end{abstract}
\maketitle

\section{Introduction}

Few-body quantum systems composed of atoms and molecules are some
of the most simple structures that constitute the building blocks
of matter. Despite this simplicity their study has recurrently been
challenging. One of the reasons being that one cannot make use of
standard statistical methods and still there are enough degrees of
freedom to make it a complex problem, often not solvable for as few
as three bodies. The interest in few body systems is manifold and
has over the time appeared in nuclear and particle physics as well
as in atomic and molecular studies. Currently, a renewed interest
also emerged in relation to the experimental study of Bose and Fermi
gases, since few body interactions may play a far from trivial role
in their behavior \cite{blume}. Moreover,the recent and impressive
development of the technology associated to the study of the Bose-Einstein
condensation phenomena in fields such as ultracold gases, Mott insulators
and optical lattices led to the possibility of controlling in an increasingly
precise way the number of atoms trapped in a well. \\

In particular, a great deal of interest has been devoted to the study
of distinguishable trapped few-fermion systems. The most recent experimental
achievement being the realization of a system of two fermionic atoms
of $^{6}\mathrm{Li}$ with tunable interactions \cite{Jochim1,Jochim2}.
In this experiment, the ground state energy of the system was measured
and compared to an analytical result that exists in this particular
case \cite{Bush} (see also \cite{Calarco}), which however is not
extendable if one includes more atoms. Therefore a good approximation
that may be generalized to more than two atoms is of interest. It
is worthwhile to mention here that the Hamiltonian employed in the
calculation of the ground-state energy of this 2-fermion experiment
is basically equivalent to the one used to discuss the existence of
exotic pairing mechanisms closely related to the elusive Fulde-Ferrell-Larkin-Ovchinnikov
(FFLO) \cite{FFLO1,FFLO2} state. In that case one deals with a higher
number of particles, in addition to a spin imbalance, and the thermodynamical
Bethe ansatz coupled to a local density approximation was used to
discuss the resulting phase diagrams and density profiles of trapped
fermionic $^{6}\mathrm{Li}$ atoms in 1D tubes \cite{Hulet,Batchelor2,Bolech}.\\

With the above motivations, viewing the prospect of new few body experiments
\cite{Jochim3}, we propose an alternative possibility, a variational
approach based on the use of the Bethe-ansatz solution for a system
with delta-function interactions. Our choice will take into account
the knowledge of the exact solution of the one dimensional many-body
system with repulsive or attractive delta-function potentials \cite{Takahashi,Yang,gaudin}
and consider the trapping as a kind of perturbation. By this we mean
that the bulk of our ansatz is supposed to grasp the behavior of the
interacting particles which happen to be trapped in a harmonic well.
To the extent of our knowledge this is an unexplored possibility and,
for that matter, one that has the potential to be systematically generalized
from two to more particles. Our approach consists in calculating the
ground state of the few-fermion model having in mind the variational
principle, such that the actual ground state energy is smaller than
the ground state of the Hamiltonian with delta interactions, which
we know exactly by the Bethe-ansatz methods, plus a part that is the
mean value of the harmonic potential for our ansatz.\\

In the following we develop our systematics for the variational calculation
of the ground state of a two-fermion system. The next section will
be devoted to set forth the system of two interacting fermions that
we are investigating, then in section 3 we introduce our variational
ansatz, which as mentioned is inspired in a paradigmatic solution
for one dimensional systems \cite{Takahashi,Yang,gaudin,Baxter},
the Bethe-ansatz \cite{Bethe}. In section 4 we present our results
for the repulsive and attractive cases and in section 5 these results
are compared with the one obtained in \cite{Bush,Calarco} for the
relative motion. In the Appendix we provide details concerning the
construction of the Bethe ansatz part in absolute coordinates and
briefly discuss its extension to the general case of $N$ fermions.

\section{System}

Let us consider a system of two interacting fermions, for instance
two fermionic atoms with mass $m$, in an axially symmetric harmonic
trap with angular frequency $\omega$. Such a system can be described
by the following Hamiltonian:

\begin{equation}
H=-\frac{\hbar^{2}}{2m}\frac{\partial^{2}}{\partial x_{1}^{2}}-\frac{\hbar^{2}}{2m}\frac{\partial^{2}}{\partial x_{2}^{2}}+V_{A}\left(x_{1},\, x_{2}\right)+V_{I}\left(x_{1}-x_{2}\right),
\end{equation}
where $x_{1}$ and $x_{2}$ denote the position of the two fermions
and $V_{A}\left(x_{1},\, x_{2}\right)$ represents the trapping potential.

\begin{equation}
V_{A}\left(x_{1},\, x_{2}\right)=\frac{1}{2}m\omega^{2}{x_{1}}^{2}+\frac{1}{2}m\omega^{2}{x_{2}}^{2}.
\end{equation}
For sufficiently low energies the interaction potential $V_{I}$ can
be taken as a delta-function contact potential, such that,

\begin{equation}
V_{I}\left(x_{1}-x_{2}\right)=2c\delta\left(x_{2}-x_{1}\right),
\end{equation}
where $c$ is the interaction strength. The potential is repulsive
or attractive, respectively for $c>0$ or $c<0$.\\

Here, as the harmonic potential and the kinetic energy are quadratic
it is convenient to separate the relative motion from the center of
mass motion. This can be easily attained by using center of mass and
relative coordinates given by, 
\begin{equation}
X=\frac{x_{1}+x_{2}}{2},\,\,\, x=x_{2}-x_{1}.
\end{equation}

One can then decompose the total Hamiltonian in the center of mass
$H_{CM}$ and relative motion $H_{rel}$ parts,

\begin{equation}
H_{CM}=-\frac{\hbar^{2}}{2M}\frac{\partial^{2}}{\partial X^{2}}+\frac{1}{2}M\omega^{2}X^{2},
\end{equation}

\begin{equation}
H_{rel}=-\frac{\hbar^{2}}{2\mu}\frac{\partial^{2}}{\partial x^{2}}+2c\delta\left(x\right)+\frac{1}{2}\mu\omega^{2}x^{2}.
\end{equation}
In the above, $\mu=\frac{m}{2}$ is the reduced mass and $M=2m$ the
total mass. It can be seen that the eigenfunctions and eigenenergies
of $H_{CM}$ are those of the harmonic oscillator. Notice that for
the general $N$ case, by the use of Jacobi coordinates, the Hamiltonian
is also separable (see Appendix).\\

Now, for the Hamiltonian $H_{rel}$, we shall apply the variational
principle,

\begin{equation}
E_{GS}\leq\frac{\left\langle \psi\right|H_{rel}\left|\psi\right\rangle }{\left\langle \psi\right|\left.\psi\right\rangle },
\end{equation}
where $\psi\left(x\right)$ is a continuous trial wavefunction.\\

The novelty in our approach is that the trial function, which we denote
$\psi\left(x,\,\left\{ \alpha,L\right\} \right)$, will encompass
the Bethe ansatz concept \cite{Takahashi,Bethe,Yang,gaudin}. The
parameter $\alpha$ controls the decay of the trial function outside
the trap and $L$ indicates the limit where this decay starts. Inside
the trap, where the contact interaction is relevant the trial function
will take the form of the Bethe ansatz. As usual a variation of these
parameters provides a minimal value, which should approximate the
ground state of the system. In that way we have a wavefunction that
gives a realistic picture of the physical processes involved.

\section{Ansatz}

In the present section we exhibit our variational ansatz. Further
details concerning the construction of the Bethe ansatz part can be
found in the Appendix, where we also briefly discuss its extension.\\

As shown in $Fig.\,1$ there are three relevant regions for a wavefunction
of our system. We can delineate these regions by the parameter $L$.
Our variational ansatz assumes then the following configuration:

\begin{equation}
\psi=\begin{cases}
\psi_{I}=e^{-\alpha\left(x+L\right)^{2}}\psi_{II}\left(-L\right) & -\infty<x<-L\\
\psi_{II}=(e^{ikL}e^{-ikx}+e^{ikx})\Theta\left(x\right)\\
\,\,\,\,\,\,\,\,\,\,\,\,+(e^{ikL}e^{ikx}+e^{-ikx})\Theta\left(-x\right) & -L<x<+L\\
\psi_{III}=e^{-\alpha\left(x-L\right)^{2}}\psi_{II}\left(+L\right) & +L<x<+\infty
\end{cases}\label{ansatz}
\end{equation}
where $\Theta$ is the Heaviside step function.

\begin{center}
\begin{figure}[H]
\begin{centering}
\includegraphics[scale=0.69]{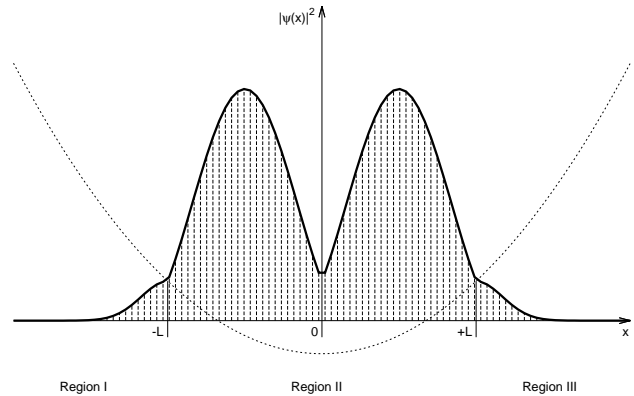} 
\par\end{centering}

\caption{Schematic representation of the normalized probability density, $|\Psi(x)|^{2}$
in the relative coordinates system. The variational parameter $L$
may be used to delineate three regions according to the boundaries
with respect to the harmonic potential.}
\end{figure}

\par\end{center}

In region $II$ $\left(-L<x<+L\right)$ the two fermion system is
subject to the contact potential and the harmonic trap. Due to the
symmetry of the system, in the vicinity of the central axis, where
$\left(x=0\right)$, the interaction term is dominant. In other words,
any contact interaction takes over the harmonic potential. For this
reason we assume that is possible to approximate the wavefunction
in region $II$ by the wavefunction that describes a system with two
distinct fermions with a contact interaction. Historically, such systems
where studied in one dimensional lattices of size \textquotedbl{}$L$\textquotedbl{}
and periodical boundary conditions, being exactly solved in \cite{Yang,gaudin}.
Later major contributions for this problem where given by \cite{Takahashi}
followed by others, such as \cite{Sutherland} and \cite{Batchelorpt,Batchelor3,he,Batchelor}.\\

Therefore, our choice for a trial wavefunction in this region corresponds
to the Bethe ansatz solution for fermions interacting trough a delta
function potential in relative coordinates, such that $\psi_{II}$
is built as the eigenfunction of the interaction Hamiltonian

\[
H_{int}\psi_{II}=\left[-\frac{\hbar^{2}}{2\mu}\frac{\partial^{2}}{\partial x^{2}}+2c\delta\left(x\right)\right]\psi_{II}=\frac{\hbar^{2}}{2\mu}k^{2}\psi_{II}.
\]

More importantly, this means that $\psi{}_{II}$ will correspond to
the eigenfunction of $H_{int}$ with energy $E_{int}=\frac{\hbar^{2}}{2\mu}k^{2}$
for all the \textquotedbl{}quasi-momenta\textquotedbl{} $k's$ that
satisfy the following equation

\begin{equation}
e^{ikL}=\frac{k+i\frac{2\mu}{\hbar^{2}}c}{k-i\frac{2\mu}{\hbar^{2}}c},\label{BAE}
\end{equation}
known as the Bethe ansatz equation.\\

Hence, for each value of the coupling $c$ we need to determine which
value of the quasi-momentum $k$ satisfies (\ref{BAE}), for the ground
state, in order to have $\psi_{II}$ completely defined.\\

A careful analysis of (\ref{BAE}) shows that the possible values
for the quasi-momenta $k$ depend on the sign of $c$ \cite{Takahashi}.
For the repulsive case $\left(c>0\right)$, only real $k$'s are ground-state
solutions of the Bethe ansatz equations (\ref{BAE}) and, accordingly,
are the values entering in $\psi_{II}$. For the attractive case $\left(c<0\right)$
the $k$'s composing the ground state are pure imaginary numbers.
We will then consider both cases separately.\\

Our proposal consists then in building the central part of our trial
function as the Bethe ansatz wavefunction for the relative motion
of two distinct fermions interacting via a delta function. This problem
is completely solvable and in its generality applied to any number
of fermions \cite{Takahashi,Yang,gaudin}. Notice that in the literature,
one usually considers the contact interaction as a perturbation to
the harmonic potential Hamiltonian. We show here how to use the full
strength of the Bethe ansatz in a variational approach.\\

Before we proceed to the analysis of the repulsive and attractive
cases we have still to explain how to deal with the continuity of
the wavefunction on the boundaries between the three regions. The
continuity condition in all the interval dictates that 
\begin{eqnarray*}
\psi_{I}\left(-L\right) & = & \psi_{II}\left(-L\right),\\
\psi_{II}\left(+L\right) & = & \psi_{III}\left(+L\right).
\end{eqnarray*}

In regions $I$ and $III$ the harmonic potential is the only one
present, so the simplest choice that take into account the system
behavior should be an eigenfunction of the harmonic oscillator Hamiltonian.
As we expect a rapid decay of the probability density in these regions,
$\psi_{I}$ and $\psi_{III}$ have the form of a Gaussian and depend
on another variational parameter, $\alpha$. It is important to notice
that although our choice is continuous for all $x$, its first derivative
is not. Later we also consider the contribution of this discontinuity
to the total value of the ground state energy.

\section{Results}

\subsection{Repulsive interaction, $\left(c>0\right)$}

In this case just some values of $k$, \textit{purely real} ones,
satisfy (\ref{BAE}) for the ground state, therefore in this subsection
we only consider $k\in\mathbb{R}$. In order to apply the variational
principle, we first compute the normalization of the wavefunction,
which yields

\begin{equation}
\left\langle \psi\right|\left.\psi\right\rangle =2\sqrt{\frac{\pi}{2\alpha}}\left[1+\cos\left(kL\right)\right]+\frac{4}{k}\left[kL+\sin\left(kL\right)\right]
\end{equation}
and then the expectation value of $H_{rel}$, which value is 
\begin{align}
 & \left\langle \psi\right|H_{rel}\left|\psi\right\rangle =\frac{\hbar^{2}\alpha}{\mu}\sqrt{\frac{\pi}{2\alpha}}\left[1+\cos\left(kL\right)\right]+\frac{2\hbar^{2}k^{2}L}{\mu}\nonumber \\
 & +\mu\omega^{2}\left[1+\cos\left(kL\right)\right]\left[\frac{1}{2}\frac{\sqrt{\pi}}{\left(2\alpha\right)^{\frac{3}{2}}}+L^{2}\sqrt{\frac{\pi}{2\alpha}}+\frac{L}{\alpha}\right]\nonumber \\
 & +\frac{\mu\omega^{2}}{3k^{3}}\left[2L^{3}k^{3}+3\left(k^{2}L^{2}-1\right)\sin\left(kL\right)+3kL\cos\left(kL\right)\right].\label{Expected_R}
\end{align}
In the expression (\ref{Expected_R}) we also considered the contribution
from the discontinuity of the wavefunction first derivative at $x=\pm L$,
that is: 
\begin{align*}
 & \lim_{\epsilon\rightarrow0}\left[\int_{-L-\epsilon}^{-L+\epsilon}\psi^{*}\left(H_{rel}\psi\right)dx+\int_{L-\epsilon}^{L+\epsilon}\psi^{*}\left(H_{rel}\psi\right)dx\right]\\
 & =-\frac{2\hbar^{2}k}{\mu}\sin\left(kL\right),
\end{align*}
which comes from the kinetic term of $H_{rel}$.

The Bethe ansatz equations (\ref{BAE}) for the ground state in the
repulsive interaction reduce then to 
\begin{equation}
k=\frac{2}{L}\arctan\left(\frac{2\mu c}{\hbar^{2}k}\right),
\end{equation}
which are much simpler to solve.\\

We have then all the necessary ingredients to proceed with the numerical
minimization of $\frac{\left\langle \psi\right|H_{rel}\left|\psi\right\rangle }{\left\langle \psi\right|\left.\psi\right\rangle }$
with respect to the parameters $\alpha$ and $L$. Basically to each
assigned $c$, we sweep over all values of $\alpha$ and $L$, calculate
$k$ for each $L$ and establish the parameters $\alpha^{*}$ and
$L^{*}$ such that $\frac{\left\langle \psi\right|H_{rel}\left|\psi\right\rangle }{\left\langle \psi\right|\left.\psi\right\rangle }$
takes the least possible value. In this way we determine the ground
state energy of the two fermion system as a function of the coupling
$c$ via the variational principle, where the trial wavefunction is
constructed by means of the Bethe ansatz. This result is depicted
in $Fig.\,2$ using the physical variables $\epsilon$ and $a_{1D}$.
We give more details in the next section where we also compare this
result with the analytical solution \cite{Bush}.\\

Limiting case: Notice that in the limit $L\rightarrow0,\, c\rightarrow0$
(harmonic oscillator) the expression (\ref{Expected_R}) reduces to

\begin{equation}
\lim_{L\rightarrow0}\frac{\left\langle \psi\right|H_{rel}\left|\psi\right\rangle }{\left\langle \psi\right|\left.\psi\right\rangle }=\frac{\hbar^{2}\alpha}{2\mu}+\frac{\mu\omega^{2}}{8\alpha}.
\end{equation}
Upon extremization of the total energy with respect to $\alpha$ in
the limit $L\rightarrow0$, the minimum value is the one for the value
$\alpha^{*}$ of the parameter

\begin{equation}
\left|\alpha^{*}\right|=\frac{\mu\omega}{2\hbar},
\end{equation}
such that 
\begin{equation}
\left.\lim_{L\rightarrow0}\frac{\left\langle \psi\right|H_{rel}\left|\psi\right\rangle }{\left\langle \psi\right|\left.\psi\right\rangle }\right|_{\alpha=\alpha^{*}}=\frac{1}{2}\hbar\omega,
\end{equation}
which, as expected, is simply the ground state energy of the harmonic
oscillator.

\subsection{Attractive interaction, $\left(c<0\right)$}

In this case only the purely imaginary values of $k$ satisfy the
Bethe ansatz equations (\ref{BAE}) for the ground state; for this
reason we will consider $k\in\mathbb{C}$. Thus, it is convenient
to define $k=ik^{\prime},\, k^{\prime}\in\mathbb{R}$. In terms of
$k^{\prime}$, we can write (\ref{BAE}) as 
\begin{equation}
e^{-k^{\prime}L}=\frac{k^{\prime}+\frac{2\mu}{\hbar^{2}}c}{k^{\prime}-\frac{2\mu}{\hbar^{2}}c},
\end{equation}
which can be solved by numerical methods.\\

In order to apply the variational principle, we first compute the
normalization of the wavefunction, which yields 
\begin{align}
 & \left\langle \psi\right|\left.\psi\right\rangle =2e^{-k^{\prime}L}\left\{ \sqrt{\frac{\pi}{2\alpha}}\left[1+\cosh\left(k^{\prime}L\right)\right]\right.\nonumber \\
 & \left.+\frac{2}{k^{\prime}}\left[Lk^{\prime}+\sinh\left(k^{\prime}L\right)\right]\right\} 
\end{align}
and then the mean value of $H_{rel}$, obtaining 
\begin{align}
 & \left\langle \psi\right|H_{rel}\left|\psi\right\rangle =\frac{\hbar^{2}\alpha}{\mu}\sqrt{\frac{\pi}{2\alpha}}e^{-k^{\prime}L}\left[1+\cosh\left(k^{\prime}L\right)\right]\nonumber \\
 & -\frac{2\hbar^{2}k^{\prime2}L}{\mu}e^{-k^{\prime}L}+\frac{\mu\omega^{2}}{3k^{\prime3}}e^{-k^{\prime}L}\left[2L^{3}k^{\prime3}\right.\nonumber \\
 & \left.+3\left(k^{\prime2}L^{2}+1\right)\sinh\left(k^{\prime}L\right)-3k^{\prime}L\cosh\left(k^{\prime}L\right)\right]\nonumber \\
 & +\mu\omega^{2}e^{-k^{\prime}L}\left[1+\cosh\left(k^{\prime}L\right)\right]\left[\frac{1}{2}\frac{\sqrt{\pi}}{\left(2\alpha\right)^{\frac{3}{2}}}+L^{2}\sqrt{\frac{\pi}{2\alpha}}+\frac{L}{\alpha}\right],\label{Expected_A}
\end{align}
where again we considered the contribution of the discontinuity of
the wavefunction derivative at the points $x=\pm L$,

\begin{align*}
 & \lim_{\epsilon\rightarrow0}\left[\int_{-L-\epsilon}^{-L+\epsilon}\psi^{*}\left(H_{rel}\psi\right)dx+\int_{L-\epsilon}^{L+\epsilon}\psi^{*}\left(H_{rel}\psi\right)dx\right]\\
 & =\frac{2\hbar^{2}k^{\prime}}{\mu}e^{-k^{\prime}L}\sinh\left(kL\right),
\end{align*}
into the expression (\ref{Expected_A}).\\

As in the attractive case we numerically minimize the mean energy
$\frac{\left\langle \psi\right|H_{rel}\left|\psi\right\rangle }{\left\langle \psi\right|\left.\psi\right\rangle }$
with respect to the parameters $\alpha$ and $L$ and determine the
ground state energy of the system as a function of the coupling $c$.
This result is depicted in $Fig.\,2$. In the next section we make
a comparison with the analytical solution obtained in \cite{Bush}.

Limiting case: Notice again that in the limit $L\rightarrow0,\, c\rightarrow0$
(harmonic oscillator)

\begin{equation}
\lim_{L\rightarrow0}\frac{\left\langle \psi\right|H_{rel}\left|\psi\right\rangle }{\left\langle \psi\right|\left.\psi\right\rangle }=\frac{\hbar^{2}\alpha}{2\mu}+\frac{\mu\omega^{2}}{8\alpha}.
\end{equation}

Upon extremization of the total energy with respect to $\alpha$ in
the limit $L\rightarrow0$, we obtain $\left|\alpha^{*}\right|=\frac{\mu\omega}{2\hbar}$,
such that we find again the ground state energy of the harmonic oscillator.

\section{Comparison}

The results obtained in the previous section by means of the Bethe
ansatz and the variational principle to find the two fermion system
ground state as a function of the coupling parameter are presented
in $Fig.\,2$ (red straight line) . For convenience and to compare
with known results we are using the following variables, 
\begin{equation}
\epsilon=\frac{E_{GS}-\frac{1}{2}\hbar\omega}{\hbar\omega}\,\,\,\,\, a_{1D}=-\frac{\hbar^{3/2}}{2c}\sqrt{\frac{\omega}{\mu}},
\end{equation}
where $\epsilon$ denotes the energy of the ground state shifted by
the zero point energy in $\hbar\omega$ units and $a_{1D}$ is the
one dimensional scattering length.\\

The analytical solution in relative coordinates for a system of two
distinct fermions interacting via a delta function and confined in
a harmonic trap was first obtained in \cite{Bush}. Basically in this
work they expanded the unknown wavefunction in a complete set of the
simple harmonic oscillator functions. Later, these results were generalized
to different geometries of the trapping potential in \cite{Calarco},
and among others in \cite{Julienne,Julienne2,Calarco2,peres,Bush2}.\\

Essentially, the following implicit equation determining the eigenenergies
of the relative motion in a one-dimensional harmonic potential was
obtained 
\begin{equation}
2a_{1D}=\frac{\Gamma\left(-\frac{\epsilon}{2}\right)}{\Gamma\left(-\frac{\epsilon}{2}+\frac{1}{2}\right)},\label{calarco}
\end{equation}
where $\Gamma\left(x\right)$ is the complete gamma function. \\

This solution for the ground state is plotted in $Fig.\,2$ (black
dotted line). We can observe a very good agreement between this result
and the result that we obtained combining the Bethe ansatz and the
variational principle (red line). This places our ansatz as a potential
candidate for the extension to more than two fermions, were an analytical
solution does not exist. The fact that the measured properties of
this system \cite{Jochim1} may, with a good agreement, be compared
with the theoretical results \cite{Bush} makes this subject even
more captivating.

\begin{center}
\begin{figure}[H]
\begin{centering}
\includegraphics[scale=0.69]{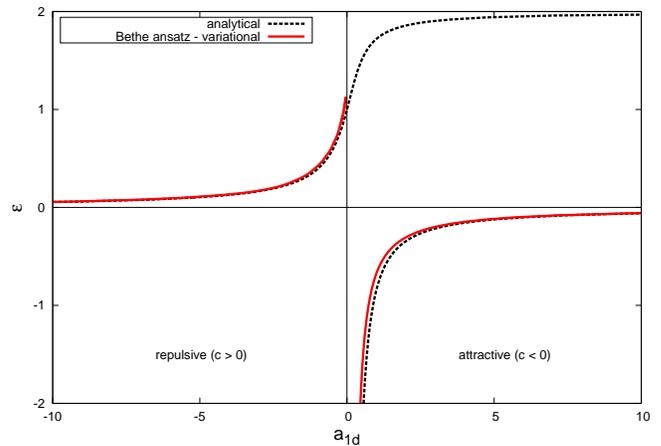} 
\par\end{centering}

\caption{ Energies for the ground state of the relative motion for two distinct
fermions interacting via a delta function potential and confined in
a harmonic trap of frequency $\omega$. The analytically obtained
energy levels (black dotted line) are compared to the results for
the energy given by the combination of the Bethe ansatz with the variational
principle (red straight line).}
\end{figure}

\par\end{center}

\section{Conclusion}

In this work we obtained the ground-state energy of two distinct fermions
in a 1D harmonic trap within a variational approach, but from a distinct
perspective, aiming a new view for the problem of few fermions. The
reasoning beneath our variational ansatz choice was to exploit the
exact solution for the one dimensional system of fermions interacting
by means of a contact potential solution, the Bethe ansatz. Usually,
in the literature, one takes a route different from ours by considering
the harmonic trap Hamiltonian as the relevant one and the contact
interaction as a perturbation. But, since for delta function interactions,
we have at our disposal the Bethe ansatz technology it is almost natural
to use it. Thus, we chose a trial wavefunction for this system that
contains a great deal of information about the physics of the two
fermions inside the trap and supplement it by the knowledge of the
harmonic oscillator Hamiltonian. The good agreement between our results
and existing analytical results shows that our ansatz fulfills our
expectation and has the potentiality to shed light on the spectrum
of strongly correlated few-body quantum systems. Using the methods
established in this work it is in principle possible to extend our
studies to more complex systems, composed of three or more fermions,
which are currently of experimental interest \cite{Jochim3}, and
also in this case one can profit of the exact solution for the contact
interaction. The procedure for higher $N$ brings however a substantial
operational growth as there are $N!\times N!$ coefficients of the
Bethe ansatz to be determined and the number of regions of the complete
variational ansatz, as in (\ref{ansatz}), increases correspondingly.

\section*{Acknowledgments}

A. Foerster thanks S. Jochim for inspiring discussions during the
BEC 2011 conference. The authors are grateful to C. C. N. Kuhn for
valuable discussions. The authors acknowledge CNPq - Conselho Nacional
de Desenvolvimento Cient\'ifico e Tecnol\'ogico for financial support,
I. Roditi also thanks FAPERJ - Funda\c{c}\~ao Carlos Chagas Filho de Amparo
\`a Pesquisa do Estado do Rio de Janeiro for financial support.

\section*{Appendix}

\global\long\def\theequation{A-\arabic{equation}}

We develop here, in detail, how we built the trial wavefunction for
the two body problem and then we indicate how to apply the same principles
for a higher number of fermions.

The rationale we use in our construction is the Bethe ansatz method
for obtaining the energy spectra of exactly solvable Hamiltonians.
Let us then consider two fermions interacting through a delta function
potential in a one dimensional system with periodicity $L$, which
has the following Hamiltonian

\begin{equation}
H=-\frac{\hbar^{2}}{2m}\frac{\partial^{2}}{\partial x_{1}^{2}}-\frac{\hbar^{2}}{2m}\frac{\partial^{2}}{\partial x_{2}^{2}}+2c\delta\left(x_{1}-x_{2}\right).\label{App1}
\end{equation}
where $x_{1}$ and $x_{2}$ are the position of each fermion and $c$
is the interaction strength. The most general wa\-ve\-fun\-ction
for such system in absolute coordinates in the region $x_{1},x_{2}\,\in\left[-L/2,+L/2\right]$
can be written as \cite{Takahashi}

\begin{widetext} 
\begin{eqnarray}
\psi\left(x_{1},x_{2}\right) & = & \left[a_{12}^{12}e^{i\left(k_{1}x_{1}+k_{2}x_{2}\right)}+a_{21}^{12}e^{i\left(k_{2}x_{1}+k_{1}x_{2}\right)}\right]\Theta\left(x_{2}-x_{1}\right)\label{App2}\\
 &  & +\left[a_{12}^{21}e^{i\left(k_{1}x_{2}+k_{2}x_{1}\right)}+a_{21}^{21}e^{i\left(k_{2}x_{2}+k_{1}x_{1}\right)}\right]\Theta\left(x_{1}-x_{2}\right).\nonumber 
\end{eqnarray}
\end{widetext} where $k_{1}$ and $k_{2}$ are the \textquotedbl{}quasi-momenta\textquotedbl{}
for the fermions and the coefficients \textquotedbl{}$a$\textquotedbl{}
are to be determined by physical arguments. The action of the Hamiltonian
on the wavefunction results in

\begin{equation}
H\psi\left(x_{1},x_{2}\right)=E\psi\left(x_{1},x_{2}\right)+undesirable\, terms,
\end{equation}
where the \textquotedbl{}undesirable terms\textquotedbl{} are functions
of $k_{1}$ and $k_{2}$. When, as usual, one requires that these
terms be null, $k_{1}$ and $k_{2}$ must satisfy certain consistency
relations known as the Bethe ansatz equations \cite{Takahashi}. These
depend on the wavefunction symmetry. Energy and momentum are given
respectively by

\begin{equation}
E=k_{1}^{2}+k_{2}^{2},\,\,\, K=k_{1}+k_{2}
\end{equation}
Once the system is in a spin singlet configuration (antissymetric)
the wavefunction must be spatially symmetric,

\begin{equation}
\psi\left(x_{2},x_{1}\right)=\psi\left(x_{1},x_{2}\right),
\end{equation}
this implies that

\begin{equation}
\begin{cases}
a_{12}^{12}=a_{12}^{21}\equiv a_{12}\\
a_{21}^{12}=a_{21}^{21}\equiv a_{21}
\end{cases}
\end{equation}
Besides, the periodic boundary conditions 
\begin{equation}
\psi\left(x_{j}=-L/2\right)=\psi\left(x_{j}=+L/2\right),\,\,\, j=1,2
\end{equation}
lead to the relations

\begin{equation}
\begin{cases}
a_{12}=a_{21}e^{ik_{1}L}\\
a_{21}=a_{12}e^{ik_{2}L}.
\end{cases}
\end{equation}

It can be shown that, for the ground state, $K=0\Rightarrow k_{2}=-k_{1}$.
Therefore, we can write the symmetric wavefunction with periodical
boundary conditions in terms of the absolute coordinates as

\begin{widetext} 
\begin{eqnarray}
\psi\left(x_{1},x_{2}\right) & = & a_{21}\left[e^{ik_{1}L}e^{ik_{1}\left(x_{1}-x_{2}\right)}+e^{-ik_{1}\left(x_{1}-x_{2}\right)}\right]\Theta\left(x_{2}-x_{1}\right)\\
 &  & +a_{21}\left[e^{ik_{1}L}e^{-ik_{1}\left(x_{1}-x_{2}\right)}+e^{ik_{1}\left(x_{1}-x_{2}\right)}\right]\Theta\left(x_{1}-x_{2}\right).\nonumber 
\end{eqnarray}
\end{widetext} Then, for convenience, defining $x=x_{1}-x_{2}$ and
$k\equiv k_{1}$, we obtain our ansatz in the relative coordinates
system: 
\begin{eqnarray}
\psi\left(x\right) & = & a_{21}\left[e^{ikL}e^{ikx}+e^{-ikx}\right]\Theta\left(-x\right)\\
 &  & +a_{21}\left[e^{ikL}e^{-ikx}+e^{ikx}\right]\Theta\left(x\right)\nonumber 
\end{eqnarray}
The constant $a_{21}$ is obtained by the normalization condition.
The above wavefunction is the eigenfunction of the interaction Hamiltonian
in relative coordinates in the $-L\leq x\leq L$ interval, and constitutes
the central part of our ansatz. In the $x\rightarrow\pm\infty$ limits
it is expected that the wavefunction exhibits an asymptotic behavior
similar to the behavior of an harmonic oscillator wavefunction with
exponential decay. Then the trial wavefunction in the other intervals
is obtained by the condition of continuity of the wavefunction on
the boundaries between all regions. \\

The generic N fermion case can be dealt with in a completely analogous
way. The wavefunction in absolute coordinates is similar to that of
Eq.(\ref{App2}) with coefficients $a_{...\lambda\mu...}^{...lm...}$
(the indices run from $1$ to $N$). It is possible then to proceed
in exactly the same way as before for all other terms. Requiring the
same physical principles as above for all the $N!$ regions of the
type $x_{1}\le x_{2}\le\,...\,\le x_{N}\,\in\left[0,L\right]$ respecting
the Bethe ansatz with periodicity $L$, a complete wavefunction in
absolute coordinates can be constructed. For instance, for the case
$N=3$ we have, in a compact form:

\begin{widetext}
\[
\psi\left(x_{1},x_{2},x_{3}\right)=\sum_{\underset{l\neq m\neq n}{l,m,n=1}}^{3}\sum_{\underset{\lambda\neq\mu\neq\nu}{\lambda,\mu,\nu=1}}^{3}a_{\lambda\mu\nu}^{lmn}e^{i\left(k_{\lambda}x_{l}+k_{\mu}x_{m}+k_{\nu}x_{n}\right)}\Theta\left(x_{n}-x_{m}\right)\Theta\left(x_{m}-x_{l}\right)
\]

\end{widetext} An important element here is that the coefficients
of the Bethe ansatz, are related through a well established transformation
where the operators are given by \cite{Takahashi,GuYang} and satisfy
the Yang-Baxter equation \cite{Yang,Baxter}. It is important to note
that, in this case, it is convenient to change from the absolute coordinates
system to Jacobian coordinates and it is possible to show that for
both the contact and trapping interaction the resulting Hamiltonian
is separable in center of mass and relative coordinates \cite{felix}.
This enable us to, knowing the Bethe ansatz result for absolute coordinates,
obtain the result for the relative coordinates Hamiltonian. In other
words the described procedure, when $N$ is increased, though cumbersome
(the number of coefficients increases as $N!^{2}$), allows one to
build the Bethe ansatz part of the whole variational ansatz. The caveat,
of course, is that the number of regions, such as in (\ref{ansatz}),
where one has to use continuity conditions also increases accordingly.

\end{document}